# A Framework for Conscious Information Processing

*Balaram Das[1]*

Department of Philosophy, University of Adelaide, South Australia 5005, Australia.

This paper exploits the fact that the variability in the inter-spike intervals, in the spike train issuing from a neuron, carries substantial information regarding the input to the neuron. A framework for neuronal information processing is proposed which utilizes the above fact to distinguish phenomenal from non-phenomenal mental representation. In the process, an explanation is offered as to what it is, in the nature of conscious mental states, that imparts them a subjective feeling – there is something it is like to be in those mental states. To give empirical support, it is shown how the proposed framework can neatly explain, the delay in eliciting conscious awareness as observed by Libet and the related backwards referral in time.

**1. Introduction**

Phenomenal experience is our first person view of the surrounding world and it is private. If you are sipping a cup of coffee, you can ask yourself – what is it like (for you) to taste the coffee, to smell its distinct aroma, to hold the cup and feel its shape and, feel the warmth of the coffee inside. It is a basic assumption that you will get some sort of answers to these questions if you are in a conscious state. You will have difficulties in expressing your feelings clearly, but you will not deny those feelings. There is something it is like to be you at any point of time when you are conscious. To paraphrase Nagel [1974], the essence of the belief that any organism, bats for example, is conscious is that there is something it is like to be that organism - there is something it is like to be a bat. This is probably the closest we can come to a definition of consciousness or phenomenal consciousness as Block [1995] would like to distinguish it.

We are also intentional beings; our mental states represent the world surrounding us. During the course of our everyday life we generate mental states with contents such as beliefs, desires, perceptions etc. All such mental states are about or refer to things in our environment. A perceptual state can refer to, or have as its content, a red apple and one may be entertaining a belief about snakes, that they are all poisonous. Such representations are ubiquitous, however theories of representation insist that the contents of these intentional states be evaluable for consistency and truth with respect to the states of the world i.e, beliefs as true or false, memories as accurate or inaccurate, perceptions as veridical or illusory – e.g. the belief that all snakes are poisonous is false.

---
[1] Email:
balaram.das@phenomenologist.org

These two central aspects of mind, phenomenal and intentional, have given rise to the representationalist thesis of consciousness, which claims that phenomenal properties are representational in character, hence if mental representation is well understood then so would be the nature of phenomenal experience. A question that arises immediately is this; are mental representations all there is to phenomenal experience? Philosophers are divided on this issue. Reductive representationalists whose main proponents are Dretske [1995] and Tye [1995] among others, would answer to the above question in the affirmative. Non-reductive representationalists [Chalmers 2004, see Pitt 2005 for a review] hold a more moderate view; according to them not all phenomenal properties of a mental state can be reduced to representational properties. The main difficulty confronting the reductive agenda is to provide an account that distinguishes between conscious and unconscious experience. Reductive representationalists endeavour to do this by appealing to the notion of functionality. For example Tye [1995] casts his thesis along the lines of *PANIC*: for Poised, Abstract, Non-conceptual, Intentional Content. Phenomenal representations according to him are *poised* to have appropriate behavioural impact. This stipulated functionality is then exploited to rule out non-phenomenal representations. This is rather problematic as Carruthers [2000, Chapter 6] points out; it is possible to instantiate functional properties without any associated phenomenology. In other words, there are mental states with representational contents which guide behaviour, but as far as the subject is concerned there is nothing 'it is like to be in these states'[2]. Moreover an important fact that remains unexplained is this: how can the mere availability of a mental state towards some functional role impart that state a subjective feeling that there is something it is like to be in that state.

In this paper I propose a manner of representation that can distinguish phenomenal from non-phenomenal representations without appealing to functionalism. I show that it is possible to differentiate between phenomenal and non-phenomenal representations by detecting the differences in the intrinsic nature of these representations - by understanding how these representations were created in the first place or in other words by gaining an insight into the nature of information processing that went into the creation of such representations. In the process I will provide a plausible explanation as to what is it in the nature some representations that imparts them a subjective feeling – there is something it is like to be in those mental states. Let me start by explicating some aspects of the information processing conducted by the central nervous system.

---

[2] Furthermore an appeal to functionalism to characterise mental states is flawed at a more fundamental level according to O'Brien and Opie [2004]. They argue that " … a theory (of mental representation) must explain mental representation in a fashion consistent with its causal role in shaping appropriate behaviour…. We shall call this the *causal constraint* on a theory of mental representation.
Remarkably, even though the well-known causal, functional, and teleosemantic theories of mental representation are all naturalistic, they *all* violate the causal constraint. Despite their internal differences, these theories ultimately treat mental content in terms of the appropriate behaviour that cognitive subjects are capable of exhibiting towards their environments. And any theory that treats mental content in terms of intelligent behaviour is in principle unable to explain how mental representation is causally responsible for such behaviour. A thing cannot cause what it is constituted by."

## 2. Hierarchies in neural information processing

*2.1 Hierarchies and feature detectors*

The main function of the sensory cortex, as Koch [2004, pp22-23] points out, "is to construct and use highly specific *feature detectors*, such as those for orientation, motion or faces". To this purpose the sensory cortex is arranged in a semi-hierarchical manner. Let us concentrate on the anatomical hierarchy of the visual system, which is better understood. The retinal ganglion cells in the retina receive the visual input from the environment and respond simply to whatever contrast is present at that point in space. Feed forward connections from the retina carry the stimulus to the low level cortical areas V1 and V2, where the signals are integrated to represent simple features such as lines or edges of specific orientation and location. From there on the visual signals propagate to higher cortical areas through further neural projections. At the next higher levels V3, V4 and medial-temporal area MT, signals are further integrated over their spatial parameters to represent more global features. Still higher levels, inferotemporal area IT, prefrontal area PF, etc., implement further integration of the visual signal to represent abstract forms, objects and categories [Hochstein and Ahissar 2002].

To see things clearly let us build a simple picture as follows. Let us say a population of neurons $W$ constitutes the projective field of a population $V$ which in turn is the projective field of a set of neurons $U$. Stimulating $U$ will stimulate $V$ which in turn will stimulate $W$; or $U, V, W$ specify an ascending hierarchy of neural groups. Neurons in $V$ receive inputs from neurons in $U$ and integrate them in some linear or non-linear fashion. Let us say a subset $\Delta$ of $V$ neurons, is primed to detect a particular significant correlation in the input from the $U$ neurons, i.e. $\Delta$ can detect some feature implicit in the activities of the $U$ neurons. Thus the activities of neurons in this subset $\Delta$ express this feature explicitly. Different subsets in $V$ would serve as different feature detectors and any such subset will modify its properties, predominantly synapses, to become a better detector of the feature it detects. Neurons in $W$ detect correlations in the activities of the neurons in $V$. Hence the features detected by the $W$ neurons are more complex as they express correlations in the features already detected by the $V$ neurons. I was careful to call the above a simple picture because things are rather more involved; one often finds feedback connections from zones higher in the hierarchy to lower ones and connections among zones at about the same level abound. Neuroscientists to a large extent, agree that as a rule (with exceptions), feedforward connections are stronger and drive the stimuli up along the hierarchy while feedback connections only serve to modify the propagation of the stimuli [Koch 2004 pp 126-127].

Thus the classical receptive fields of neurons higher in the (visual) hierarchy are larger and correspond to more complex features than those that are lower down [Crick and Koch 2003]. What is more, as one ascends up along the hierarchy, not only do neurons become increasingly selective for more and more complex features, but also the code becomes increasingly *sparse*. In other words up along the hierarchy fewer and fewer neurons get activated by a given stimulus. Sparse coding, according to Olhausen and Field [2004], confers several advantages – among others, it makes the structure in natural signals explicit, and it represents complex data in a way that is easier to read out at subsequent levels of processing.

The ability to detect specific features enables a subject to react to his or her environment rapidly and accurately. For example there are nearly 200 million photoreceptors in the eyes collecting visual information. This sea of data would be a hindrance to a subject facing a threat from the environment. What the subject needs is to extract the specific features that represent the threat, from the impinging data, and react to it aptly. This is what feature extraction in terms of sparse coding does; it makes the threatening features explicit in a way that makes it easier for subsequent levels of processing to assess and react to the threat appropriately. In the process a lot of data that are irrelevant to the perception of the threat are suppressed or made unavailable to the process of threat assessment.

*2.2 Hierarchies and conscious perception*

Representations at higher levels are less particular about spatial features such as precise location, viewpoint, lighting and colour, instead neural processing at higher levels produce representations that are comparatively more global and abstract. In other words neural activities at higher levels generalize over space, size, viewpoint etc. and thereby capture the presence of basic categories or object types but not their precise parameters. Hence what is represented at higher levels of the visual path is a *gist of the scene with blindness to the details* [Hochstein and Ahissar 2002].

How does the above picture help explain conscious perception? Crick and Koch [2003] argue that, given a visual input, the neural activity first travels rapidly and unconsciously up the visual hierarchy to a high level where a gist of the scene is produced. Neural activities then start moving backwards down the hierarchy producing conscious perception along the way. The first stages to reach consciousness are therefore the higher levels followed by the lower levels which add details to the downwards moving conscious activation. Hochstein and Ahissar [2002] provide a similar explanation, with more details, which they call *the reverse hierarchy theory*. According to them: "Processing along the feedforward hierarchy of areas, leading to increasingly complex representations, is automatic and implicit, while conscious perception begins at the hierarchy's top, gradually turning downward as needed. Thus, our initial conscious percept—*vision at a glance*—matches a high-level, generalized, categorical scene interpretation, identifying 'forest before trees'. For later *vision with scrutiny*, reverse hierarchy routines focus attention to specific, active, low-level units, incorporating into conscious perception detailed information available there. …The large receptive fields of high cortical areas are reflected in the spread attention of initial perception, while smaller low-area receptive fields are responsible for later focused attention."

Clearly what the above authors are emphasising is focussing of attention rather than the feeling of phenomenal experience. Yes, if attention to details at successive levels is what needs explaining then a downward flowing information processing, against the hierarchies, is an explanation. However conscious perception is not the same as attention to details. What needs explanation is this – if attention is focussed on a certain level, how does the subject become conscious of the features extracted at that level? I propose that a subject consciously perceives the features extracted at a certain level when she has a sense of perceiving these features within the context of yet more elementary features extracted at lower levels without actually shifting her attention to those lower levels. To this purpose let me explore a particular way in which neurons could be processing information.

# 3. Conscious information processing

*3.1 Information coded in inter-spike intervals*

Neurons process information by integrating a large number of input signals and generating a train of discrete action potentials or spikes as output. These spikes are usually generated at irregular time intervals. A question that immediately arises is whether this variability in inter-spike intervals (ISI) carries information or reflects noisy neural mechanisms. Here I argue that variability in ISI does carry information; information that is vital for conscious perception.

*In vitro* experiments with neural tissues suggest that neurons, when subjected to stimuli with transients resembling synaptic activity, can generate spikes with very fine temporal precision [Manien and Sejnowski 1995]. In other words when a stimulus is repeated the spiking pattern in the spike train is also repeated spike for spike in a highly reliable fashion. This finding is somewhat supported by recent *in vivo* experiments [see Bohte 2003 for a review]. For example, Reinagel and Reid [2000] show that lateral geniculate nucleus (LGN) relay cells can be remarkably deterministic, with reliable and temporally precise responses controlled primarily by the visual stimulus and altered very little by noise. These findings suggest that neurons in their natural surrounding can be effectively modelled as deterministic mechanisms. Furthermore environmental stimuli, by the time it reaches higher cortical levels, would have undergone a series of integrative processes by such neurons acting deterministically. Hence, for the purpose of tractability, one can make a plausible assumption that the input to cortical neurons at higher levels can be modelled to have deterministic albeit chaotic structure. In other words the input to a cortical neuron can be assumed to be generated by a chaotic dynamical system – its behaviour is neither purely random nor perfectly predictable.

The question I now pose is the following. If the input to a cortical neuron is taken to be generated by a chaotic dynamical system Đ, is it possible to gain knowledge about this dynamical system by analysing the train of spikes that issues from the neuron? Note that the train of spikes can be interpreted as the outcome of a series of measurements conducted by the neuron on Đ over time. What I am asking is this – can we utilise this series of scalar measurements to conduct an analysis that provides an understanding of the multidimensional character of the dynamical system Đ. In a seminal paper Sauer [1994] answered the above question in the affirmative, demonstrating in the process that the variability in spike generation carries information regarding the input to the neuron. Let me explain briefly.

Let $\mathbb{R}^k$ be the state space of the dynamical system Đ. The time evolution of the dynamical system is completely specified by tracing the trajectory of the state point $S(t)$, say, in the state space. Let us say that the trajectories of our dynamical system, from various initial conditions, are attracted to a $d$-dimensional attractor $A$ in the state space. With Đ representing the input applied to the neuron, let us say we observe a train of spikes at firing times $t_0 < t_1 < t_2 < \ldots$, etc. From these firing times we can construct the series of inter-spike intervals (ISI) defined as $\tau_i = t_{i+1} - t_i$ for $i = 0, 1, 2 \ldots$. Using these inter spike intervals let us construct the set of $m$-tuple ISI vectors $\{\tau_i, \tau_{i+1}, \ldots, \tau_{i+m-1}\}$, for $i = 0, 1, 2 \ldots$, etc. Each of these vectors marks a point in $\mathbb{R}^m$, the set of all these points can be assigned a topology to form a topological space $B$, say, in $\mathbb{R}^m$. Assuming that the firings of the neuron follow an integrate-and-fire process, Sauer [1994] showed that under certain genericity conditions on the underlying dynamics of the input signal and firing threshold, there is a one-to-one correspondence between the attractor $A \subset \mathbb{R}^k$ and the space $B \subset \mathbb{R}^m$, if $m > 2d$. In other words there exists a one-to-one mapping $f : B \to A$ where $f(\{\tau_i, \tau_{i+1}, …, \tau_{i+m-1}\}) = S(t_i)$ for $i = 0, 1, 2 …$; $f$ preserves the topology though not necessarily the geometry. I will refer to $B$ in $\mathbb{R}^m$ as a faithful (topologically) reconstruction of the attractor $A$ in $\mathbb{R}^k$.

The above result has also been confirmed using a variety of neuron models, including leaky integrate-and-fire models and variations on the FitzHugh-Nagumo model [Racicot and Longtin 1997]. Furthermore, Richardson et. al. [1998] have extended this work to *in vitro* studies on rat cutaneous afferents. Their investigation shows that when such a neuron is driven by a chaotic system the resulting ISI series can be used to reconstruct the chaotic attractor for predictive purposes.

As noted earlier the train of spikes can be interpreted as the outcome of a series of measurements conducted by the neuron on Ð; or what the neuron is essentially doing is recording the evolution of a property of the dynamical system Ð. This property is most probably a macroscopic property of Ð composed of its microscopic parameters like its generalised degrees of freedom. The value of this macroscopic property recorded by the neuron is given by $\tau_i$ at time $t_i$. The dynamical system would have a number of degrees of freedom. Reconstructing the attractor using the series of measurements of this one property means that the evolutions of all the degrees of freedom of the dynamical system have been reconstructed. The evolution of the measured property is now embedded in the reconstructed evolution of the whole system. Here by 'embedded' I mean the following. Let $h$ be a projection $h : B \to \mathbb{R}$ given by $h(\{\tau_i, \tau_{i+1}, …, \tau_{i+m-1}\}) = \tau_i$. Then as one traces the trajectory of Ð along the proxy attractor $B$ the projection $h$ traces the evolution of the measured property.

Although it is possible to recreate the dynamics of the whole system by measuring only a single property of the system, much quantitative information is lost. The recreated attractor does not always preserve the geometry of the original attractor, only its topology is preserved. For example the distance between two points in the original attractor is not the same as the distance between the corresponding two points in the reconstructed attractor, but the neighbourhood of a point in the original attractor is mapped to a neighbourhood of the corresponding point. Hence all quantitative properties that are metric dependent are not preserved, but most qualitative properties are preserved. Nonetheless it is a remarkable fact that a faithful image of the attractor characterizing the evolution of the input dynamical system can be reconstructed from the spike train. This gives us great analytical powers; for we can analyse this proxy reconstructed attractor to draw conclusions about the topological and sometimes geometrical characteristics of the input dynamical system itself. I will take full advantage of this fact in investigating conscious information processing.

In the above discussion we considered information processed by a single neuron. What about population coding; whereby information is processed across a population of neurons? Consider a population $P$ of cortical neurons. For the sake of definiteness let us suppose that this population is arranged like a feed forward artificial neural network, i.e., there is a layer of neurons through which information is fed into $P$, and then there are intermediate layers through which the input information gets processed, and finally a layer of output neurons project the processed information to other parts of the cortex.

To facilitate later analyses, I want to look at the problem as follows. Let us say that the output layer of *P* contains *n* neurons. Stimuli received by *P* through the input layers get processed by the internal layers, and let us denote these processed stimuli by $I_P$. These processed stimuli $I_P$ now act as input to the outer layer of *n* neurons, which react to this input by issuing a set of *n* spike trains. Equivalently, as indicated by previous analyses, I can consider the input $I_P$ to be represented by a chaotic dynamical system $Đ_P$, upon which the *n* neurons in the output layer are conducting *n* sets of measurements and issuing forth *n* spike trains as a result. The characteristics of $Đ_P$ will be determined by the nature of the stimuli input into *P*, and by the connectivity, synaptic strengths and other peculiarities of the neurons across the internal layers in *P*. Can we capture the characteristics of $Đ_P$ by analysing the *n* output spike trains?

It is possible to reconstruct the attractor of $Đ_P$ from the inter-spike intervals calculated for each of the *n* spike trains. There are no conceptual problems here but serious technical ones [Kantz and Schreiber 2003, section 9.5.2]. For our present purposes let us assume that such an attractor can be constructed. Again the *n* output neurons are essentially recording the evolutions of *n* properties of the dynamical system $Đ_P$. These properties are functions of the generalised degrees of freedom of the dynamical system – or these properties are complex features of the dynamical system composed of simpler features like the generalised degrees of freedom. The dynamical system may have more than *n* degrees of freedom. Reconstructing the attractor using these *n* series of measurements means that the evolutions of all the degrees of the dynamical system have been reconstructed. The evolutions of the *n* measured properties are now embedded in the reconstructed evolution of the whole system.

*3.2 What is it like to be in a mental state?*

As discussed in section 2.1, when a new visual scene is presented, the elementary object and environmental features are detected at lower levels of visual hierarchy. As the signals traverse up along the hierarchy these elementary features are successively integrated to form more complex features. At higher levels therefore, the features extracted are increasingly more complex and global, culminating at levels where only a gist of the scene with blindness to its details is captured.

Consider the situation where the subject is attending to some details in the visual scene. Let us say that there is a population *P* of neurons whose activities extract the features that the subject is attending to. The input to the population *P* of neurons contains the information regarding the features that are extracted at *P*. As this input is coming from the previous lower level in the hierarchy, the information it contains are in terms of features that are extracted at that lower level, and hence are more elementary than the features extracted at *P*. These more elementary features are processed by the internal layers of *P* and presented to the output layer as the input $I_P$. In other words $I_P$ captures information about the sub-features extracted at the previous lower level which have been further processed by the internal layers of *P*. We shall say that $I_P$ represents the information regarding the *sub-features processed at P*. The features extracted at *P* are composed of all or some of these sub-features processed at *P*. With time, the features extracted at *P*, would be evolving and would be perceived as snapshots at discrete time intervals. The collection of these snapshots provides the perception of the time evolutions of these features.

Again as discussed in section 3.1, we can view the input $I_P$ to be represented by a possibly chaotic dynamical system $Đ_P$ upon which the neurons in the output layer of $P$ are conducting $n$ sets of measurements and issuing forth $n$ spike trains as a result; where $n$ is the number of neurons in the output layer of $P$. These $n$ spike trains capture the evolving features that are being attended to and using the $n$ spike trains it is possible to construct a space $C$, say, which is a faithful image of the attractor of $Đ_P$. Furthermore we can consider the sub-features processed at $P$ to define a set of generalised degrees of freedom of the dynamical system $Đ_P$. Thus reconstructing the attractor of $Đ_P$, is nothing but reconstructing the evolutions of these sub-features. These evolutions are now captured in the proxy attractor $C$. The evolutions of the features extracted at $P$ are embedded in the evolutions of the sub-features processed at $P$, or more specifically, embedded in $C$.

To give an example let us imagine that our subject is seated on a park bench enjoying the scene before her – let us say she is attending to the gist of the scene before her with blindness to its details. She sees people either seated or moving around, without recognising them, she sees trees, lawns and other prominent park features without paying attention to their details. These categorical features are extracted by a population of neurons high up in the visual hierarchy. The extraction of these high level features can be used to reconstruct the attractor of the input dynamical system where the evolutions of these high level features would be embedded in the evolutions of the comparatively more elementary sub-features. The question now is this – what does such a reconstruction mean.

Let us say that there exists a set $\Pi$ of neurons which collects the features extracted by $P$ and constructs the proxy attractor $C$. From the discussion above it is clear that $C$ provides a holistic picture of the evolution of the dynamical system $Đ_P$ in terms of the evolutions of the sub-features processed at $P$, and the evolutions of the features extracted at $P$ are now embedded in $C$. Hence the activities of the $\Pi$ neurons enable the subject to perceive the evolutions of the features extracted at $P$ as embedded in the evolutions of the corresponding sub-features. It is this manner of perception, as I stipulate below, that provides the subjective feeling: there is something it is like to perceive the said features. Note that the subject is attending to the features extracted at $P$ and not to the sub-features. She is not attending to the sub-features extracted one level below. Yet she has a sense of perceiving the features within the context of all the sub-features processed at $P$; or within a holistic context.

I postulate that:
  i.   for a subject to perceive the features extracted by a population $P$ of neurons consciously, it is necessary that a reconstruction of the attractor of the dynamical system $Đ_P$ is made from the extracted features;
  ii.  when such an attractor is reconstructed, the reconstruction which captures the evolutions of the features attended to, embedded in the evolutions of the sub-features processed at $P$, imparts the subjective feeling – *there is something it is like* to attend to the features in question.

In other words, just the fact that there exists a population of neurons which extract the evolutions of certain features from the visual input does not give rise to conscious perception of those features. For conscious perception, it is necessary that the evolutions of the extracted features be represented as embedded in the evolution of a broader context.

*3.3 Neural mechanism for conscious information processing*

The above discussion indicates that for a population *P* of neurons to instantiate a conscious mental representation the following is necessary. From its input, *P* extracts the features that give rise to the contents of the mental representation. In addition there exists a subset *Π* of neurons in *P*, which uses the information about the evolutions of the extracted features to reconstruct the attractor of the dynamical system Đ$_P$. The output from *P* thus consists of two streams. The first stream consists of a set of spike trains which encodes the evolutions of the features extracted, and the second stream consists of another set of spike trains which captures the evolutions of the features extracted as embedded in the evolutions of the sub-features processed at *P*.

It is the interaction between the two streams that is responsible for generating conscious mental representation. Let me explain. Once the first stream of spiking activities have continued long enough to enable the construction of the ISI vectors of appropriate dimension the reconstruction of the attractor starts. Thus the reconstruction of the attractor and the extraction of the feature evolutions proceed hand in hand the former following the latter evolutions with a time lag. A detailed explanation of this time lag is given in section 4 below. As said before the reconstructed attractor captures the evolutions of all the degrees of freedom of the dynamical system Đ$_P$ and as one traces the trajectory of Đ$_P$ along the proxy attractor an appropriate projection traces the evolution of one or the other of the extracted features. Now it is possible to make any arbitrary projection, and an arbitrary projection would not necessarily correspond to the evolution of any of the extracted features. What informs the perceptual system to make the right projections? It is the first stream of output from *P* that carries the necessary information. Thus conscious perception is generated by the interaction of the two streams of output from *P*. The second stream which captures the evolution of all the degrees of freedom of the dynamical system Đ$_P$ provides the holistic context, while the first stream indicates the appropriate projections to be made so that the evolution of the extracted features are perceived as embedded in the evolutions of the sub-features processed at *P*[3].

Thus a population *P* of neurons that instantiates a conscious mental state may not be confined to a single cortical area. There is nothing new in this statement, neuroscientists dealing with consciousness agree unanimously on this point [see for example Crick and Koch 2003, Rees 2004]. As far as vision is concerned, it is now a rather firmly established point of view that there are two distinct neural pathways for visual perception. From the primary visual cortex V1, a *ventral* stream projects to the inferior temporal cortex, and a *dorsal* stream, to the posterior parietal cortex. As Milner and Goodale point out [Milner and Goodale 1995, Goodale 2004], the ventral pathway is necessary for delivering our conscious precepts of the world surrounding us. In contrast, the processing carried out by the dorsal stream mediates the visual control of skilled actions. The visual information that is processed by the dorsal stream for the visuomotor control is never accessible to conscious scrutiny. In the least, therefore, *P* would be composed of two sub-populations one located in the ventral pathway and the other in the dorsal pathway. The sub-population in the dorsal pathway would predominantly be responsible for extracting the features that give rise to the contents of the mental state. On the other hand the predominant function of the sub-population in the ventral pathway would be to reconstruct the attractor of the dynamical system $Ð_P$, in order to capture the evolutions of the extracted features as embedded in the evolutions of the sub-features processed at *P*. Of course, in order for the framework developed here to work, the sub-populations of *P* have to interact with each other to provide the unified sense of conscious perception, a point again supported by research in neuroscience. For example Rees [2004] writes "Reciprocal interaction between parietal and ventral visual cortex can serve to selectively integrate internal representations of visual events in the broader behavioural context in which they occur. Such network interactions may account for the richness of our conscious experience and may provide a fundamental neural substrate for visual consciousness".

---

[3] At this point one may advance an interpretation which goes as follows. The first stream of output from *P* instantiates a mental state and the second stream instantiates a higher order mental state that generates conscious awareness of the first mental state. Such a proposal would lead to *higher-order theories of consciousness* [see Carruthers 2001 for a review]. Most likely, it is possible to develop such a proposal consistently, nonetheless I do not see enough compelling reasons to do so at this point of development.

One point that is implicit in the above discussion and is worth pointing out explicitly is that the perceptual system does not have direct access to the attractor of dynamical system $Ð_P$. Its access is via the reconstructed proxy attractor. Hence its interpretations of the evolutions of the extracted features as embedded in the evolutions of the sub-features processed at *P*, are determined by the characteristics of the proxy attractor. Another important point that needs mentioning is that reconstructing the attractor of a dynamical system by a population of neurons may become infeasible if the dynamical system has a large number of degrees of freedom. This is where sparse coding comes to the rescue. As one ascends up along the hierarchy, the neural code becomes increasingly *sparse* i.e., up the hierarchy fewer and fewer neurons get activated by a given stimulus. Thus the number of degrees of freedom of the dynamical system that represents the input to a level in the hierarchy decreases successively up along the hierarchy. At higher levels therefore $Ð_P$ becomes low dimensional making it feasible to reconstruct its attractor. Since constructing such an attractor is essentially what gives rise to the subjective feeling of phenomenal experience, it is inevitable according to the framework developed here that neural correlates of visual consciousness be found at higher levels of visual processing [Rees et. al. 2002].

*3.4 Phenomenal vs. non-phenomenal representation*

With the insight gained from the above analysis it is now possible to answer the central question posed at the outset – what distinguishes phenomenal from non-phenomenal mental representations? For a mental representation (of a visual scene) to be phenomenally conscious, it is necessary that the contents of the representation capture the evolutions of the relevant features (of the visual scene) as embedded in the evolutions of the sub-features of which these features are composed of. If only the evolutions of the features are captured and not the embedding, then the mental representation remains unconscious.

**4. Libet's half-a-second delay in eliciting consciousness**

In a series of experiments that has been unique in neuroscience, Libet had the opportunity to study awake and responsive patients, whose somatosensory cortex was exposed for applying electrical stimuli for therapeutic purposes. From these experiments Libet concluded that, half-a-second continuous neural activity is required to elicit consciousness in a subject [Libet 2004, Chapter 2; Blackmore 2003, pp 57-60; see also Marchetti 2005]. I will summarise here only those of Libet's findings that I want to deal with to provide empirical support for the theoretical framework developed in the previous sections.

By stimulating the patient's somatosensory cortex with a train of electrical pulses Libet found that there is a minimum intensity for these pulses below which no sensation can be elicited. At the liminal intensity, the patients reported a feeling of sensation, coming from the related part of the skin, if the train of pulses continued for at least 500 milliseconds. In other words consciousness takes half-a-second to build up. Moreover this minimum requirement of a 500 milliseconds train of pulses to elicit conscious response was found to be independent of the frequency of the pulses. Libet also found that shorter train durations could elicit reportable sensation but for this the required intensity of the pulses would have to be very high and go into the ranges not often encountered in a person's normal everyday level of sensory experience. Libet therefore concluded that a 500 millisecond continuous neuronal activity is a minimum necessity, the *neural adequacy*, before a patient can report conscious sensation, and hence the observed half-a-second delay in eliciting consciousness.

Would such a neural adequacy be required for a reportable sensation if the patients were stimulated at the skin instead? In other words if one touches a subject, would this touch have to generate continuous neural activity for 500 milliseconds in order for the touch to be felt. Libet found that to be the case. He stimulated the skin by applying a single pulse and followed it by giving stimulation to the cortex. When the cortical stimulus was applied between 200 and 500 milliseconds after the skin stimulus the skin stimulus was not felt or the cortical stimulation retroactively masked the skin stimulation. This way Libet indirectly showed that the skin stimulus would need to generate neural activities for at least 500 milliseconds at the somatosensory cortex before it can be felt. Masking the skin stimulus before such a neural adequacy is reached will result in no reportable sensation.

However half-a-second is comparatively a long interval of time in one's perceptual activities, how come one never notices this delay. In Libet's words [Libet 1999] "We solved this paradox with a hypothesis for 'backward referral' of subjective experience to the time of the first cortical response, the primary evoked potential. This was tested and confirmed experimentally [Libet et. al. 1979], a thrilling result". The primary evoked potentials occur after about 10-20 milliseconds after peripheral stimulation. According to Libet, the time of the occurrence of the primary evoked potential acts as an epoch to which a sensation of touch is subjectively referred back to once the neural adequacy for the touch has been achieved. Note that a primary evoked potential is not generated when a train of pulses is applied directly to the somatosensory cortex. Thus, in such a case, there does not exist an identifiable epoch to which the subjective experience can be referred back to.

Let me now explain how the framework developed in Section 3 for conscious perception can neatly explain, the half-a-second delay in eliciting conscious awareness as well as the related backwards referral in time. Consider a somatosensory neuron which has been activated either by applying a pulse to the skin or by a train of electrical pulses applied directly to the cortex. As a result the neuron issues forth a train of spikes. In section 3.2, I postulated that for conscious perception of the stimulus it is necessary to reconstruct the attractor of the dynamical system representing the input to the neuron, and in section 3.1, I showed how this attractor can be reconstructed from the inter-spike intervals. To recapitulate briefly, let the time evolution of the input dynamical system be specified by the trajectory of its state point $S(t)$ and let the trajectory be attracted to a $d$-dimensional attractor $A$ in the state space $\mathbb{R}^k$. Let the neuron start firing at the time $t_0$, issuing a train of spikes at firing times $t_0 < t_1 < t_2 < \ldots$, etc. From these firing times we can construct the inter-spike intervals $\tau_i = t_{i+1} - t_i$. Using the ISIs we construct a set of *m*-tuple ISI vectors $\{\tau_i, \tau_{i+1}, \ldots, \tau_{i+m-1}\}$, for $i = 0, 1, 2 \ldots$, etc. All these vectors now define a topological space *B* in some $\mathbb{R}^m$, and if $m > 2d$, then *B* is a reconstruction of *A*. The fact that *B* is a reconstruction of *A* is indicated by the fact that one can find a one-to-one mapping $f: B \to A$ which preserves the topology.

Let us unravel this reconstruction of the attractor step by step. The first *m*-tuple ISI vector is $\{\tau_0, \tau_1, \ldots, \tau_{m-1}\}$. As $\tau_{m-1} = t_m - t_{m-1}$, this vector cannot be constructed until the spike at $t_m$ has appeared. Similarly the second ISI vector $\{\tau_1, \ldots, \tau_m\}$ cannot be constructed until the spike at $t_{m+1}$ has appeared and so on. In other words the first point of the manifold *B* will be established at time $t_m$ and the second point at time $t_{m+1}$ and so on. Therefore the reconstruction of *B*, which is a proxy of the attractor *A*, does not start at time $t_0$ but is delayed until time $t_m$. Now according to the framework espoused here the reconstructed attractor *B* is necessary for conscious perception, hence the delay in eliciting conscious sensation in Libet's experiments.

Again the reconstructed manifold *B* informs the nervous system about the input to the neuron under consideration. The nervous system does not have direct access to the attractor *A* of the dynamical system representing the input. It interprets each point in *B* to be a point in *A* and this correspondence is given by the one-to-one mapping $f: B \to A$. How does this correspondence work? Consider the *i*th point in the construction of *B* which is the ISI vector $\{\tau_i, \tau_{i+1}, \ldots, \tau_{i+m-1}\}$. Now $\tau_i = t_{i+1} - t_i$ and $\tau_{i+m-1} = t_{i+m} - t_{i+m-1}$. Hence the construction of the *i*th point in *B* depends upon the spiking times bounded by $[t_i, t_{i+m}]$; or equivalently it depends upon a series of measurements conducted by the neuron on the input dynamical system between times $[t_i, t_{i+m}]$. Hence the *i*th point $\{\tau_i, \tau_{i+1}, \ldots, \tau_{i+m-1}\}$ in *B* can be mapped onto either of the following points in *A*: either onto $S(t_i)$ or onto $S(t_{i+m})$. In other words we have two one-to-one mappings

   $f_1: B \to A$ where $f_1(\{\tau_i, \tau_{i+1}, \ldots, \tau_{i+m-1}\}) = S(t_i)$ for $i = 0, 1, 2 \ldots$, or
   $f_2: B \to A$ where $f_2(\{\tau_i, \tau_{i+1}, \ldots, \tau_{i+m-1}\}) = S(t_{i+m})$ for $i = 0, 1, 2 \ldots$.

If the points in *B* are interpreted to be points in *A* through the mapping $f_1$ then the first point in *B* which is established at time $t_m$ will correspond to the point $S(t_0)$ in *A* i.e. correspond to the state of the input dynamical system at time $t_0$. This will give rise to a backward referral in time. If on the other hand, the interpretation takes place via the mapping $f_2$, then the first point in *B* will correspond to the point $S(t_m)$ with no backward referral in time. The nervous system is able to take advantage of both these interpretation. When the input signal gives rise to a primary evoked potential, which would appear a time $t_0$, the correspondence $B \to A$ is via $f_1$, and in the absence of any primary evoked potential the correspondence is implemented through $f_2$.

## 5. Conclusion

In this paper I have exploited a few established facts in neuroscience to analyse aspects of information processing carried out by a population of neurons. Adopting knowledge from neuroscience as a basis, it has been possible to invoke the methods of non-linear analysis to understand the intrinsic nature of conscious information processing and thus get some sort of a grip over the elusive concept of phenomenal experience. The frame work proposed here provides an example of how experimentally based hypotheses, generated through analytical methods, can in turn be validated using existing empirical results. Over the years, neuroscientists employing various imaging techniques, have uncovered a plethora of facts unravelling the peculiarities of the central nervous system. It is tempting to think that analytical methods developed in non-linear sciences and information theory etc., can be employed to take advantage of this vast store of knowledge to improve upon the suggested frame work.

I wish to thank Jon Opie for his comments which clarified many points and improved the manuscript.